\documentclass[12pt]{article}
\usepackage{graphicx}
\def\address#1{\begin{center}{\it #1}\end{center}}
 \newtheorem{theorem}{Theorem}[section]

\newtheorem{remark}{Remark}[section]
\newcommand{\text}{\mbox}
\begin{document}

\title{Understanding singularities in Cartan's and NSF geometric structures.}
\author{D. M. Forni \and M. S. Iriondo \and C. N. Kozameh \and M. F. 
Parisi\footnote{E-mail: \{forni, mirta, kozameh, 
fparisi\}@fis.uncor.edu
}}

\date{}
\maketitle

\address{FaMAF, Universidad Nacional de C\'ordoba, \\
5000, C\'ordoba, Argentina }

\vspace{.2in}
\begin{abstract}
In this work we establish a relationship between Cartan's geometric approach
to third order ODEs and the 3-dim Null Surface Formulation (NSF). We then 
generalize both
constructions to allow for caustics and singularities that  necessarily
arise in these formalisms.
\end{abstract}


\section{Introduction}

During the first half of the XXth century, while trying to understand the
group of transformations of differential equations, Cartan   laid down the
foundations of modern differential geometry and established a link between
analysis and geometry. One particular example that will be discussed in this
work shows Cartan's approach to the classification of solutions of ODEs 
\cite{cartan,cartan2, chern}.
Consider the 2-dim space $E^{2}$ with local coordinates $(x,y)$ and the
following third order ODE

\begin{equation}
y^{\prime \prime \prime }=F\,(x,y,y^{\prime },y^{\prime \prime }),
\label{one}
\end{equation}
with ``prime'' denoting the derivative with respect to the independent
variable. If one performs a coordinate transformation on $E^{2}$ one gets
another ODE that is trivially related to the above equation. Cartan thus
considered the issue of how to classify solutions of third order ODEs into
equivalence classes, with two solutions belonging to the same class if the
corresponding ODEs were related by a coordinate transformation on $E^{2}$.
It is clear that one can spend many hours before finding the appropriate
coordinate transformation that will turn one ODE into the other. Cartan
showed that with the general solution of a given third order ODE, like the
above, one can explicitly construct a connection one form on a 4-dim space $%
E^{4}$ with local coordinates $(x,y,y^{\prime },y^{\prime \prime })$ (the
details are presented in the next section). Furthermore, Cartan
showed that two third order ODEs are equivalent if their corresponding
solutions yield the same geometric structure on $E^{4}$. Cartan also showed
that when $F$ satisfies a given PDE on $E^{4}$, symbolically written as $%
M[F]=0$, then the connection is torsion free and a unique conformal
structure can be given on the solution space (3-dim parameter space) of the
starting ODE.

It is worth mentioning that the same equation $M[F]=0$ arises in the three
dimensional version of the so called Null Surface Formulation (NSF) of
General Relativity (GR) \cite{2+1}. As we will see below, this is not mere
coincidence since we will show that the 3-dim version of NSF is almost
contained in one of Cartan's works \cite{cartan}. Finding the correspondence
between these two constructions was one of the motivations for writing this
work.

In the NSF the main variable is a function $u=Z(\phi )$, with $\phi $ $\in $ $%
S^{1}$, subject to a third-order ordinary differential equation of the form 
\begin{equation}
u^{\prime \prime \prime }=\Lambda \,(\phi ,u,u^{\prime },u^{\prime \prime }),
\label{ec1}
\end{equation}
where the function $\Lambda \,$ is restricted to satisfy the ``metricity
condition'' $M[\Lambda \,]=0$. The general solution of (\ref{ec1}) is given
by $u=Z(x^{a},\phi )$, with the constants of integration $x^{a}$ taken as
coordinates of the 3D solution manifold $E^{3}$. From this solution a space-time
metric for $E^{3}$ can be constructed such that the level surfaces of $Z$
are null foliations of $E^{3}$ for each fixed value of $\phi $. Furthermore,
the covector $l_{a}\equiv \partial _{a}Z$ spans, for fixed values of $x^{a}$%
, the circle of null directions and thus the parameter $\phi $ $\in S^{1}$.
Except for the relabelling of functions and the topology of the starting
space $E^{2}$, both formulations share many geometric features, with Cartan
emphasizing the role of the connection one-form and the NSF the explicit
construction of a metric on the solution space of (\ref{ec1}).

But a key point arises by noting that, by assumption, both formulations are
local constructions in the 2-dim manifold. It follows from this fact that
the solutions to (\ref{one}) are families of smooth curves and (as we will
see below) that the level sets $Z(x^{a},\phi )=const$ are smooth null
surfaces of the induced metric in the 3-dim solution space. However, we know
that characteristic surfaces develop caustics and singularities as a result
of the `focusing' properties of null geodesic congruences established in GR
singularity theorems \cite{hawking}. Moreover, the solutions to (\ref{one})
that yield these null surfaces with caustics cannot be smooth curves on $%
E^{2}$, so they must develop self-intersections and singular points.

A second motivation for this work is thus to consider the non-diffeomorphic
generalization of these geometric constructions in order to account for the
folds and non-differentiable edges of curves on $E^{2}$ that yield null
congruences with caustics. The first step is to introduce the curves on $%
E^{2}$ according to Arnold's theory of lagrangian manifolds as
\cite{arnold}: 
\begin{eqnarray}
u &=&G(p)-p\text{ }G^{\prime }(p),  \label{lagr1} \\
\phi &=&-G^{\prime }(p);  \label{lagr2}
\end{eqnarray}
($G(p)$ is called the generating function). Although $G$ may be smooth, the
curve $(u(p)$, $\phi (p))$ might have self-intersections and singular points
if $\phi (p)$ is not injective. The question is how to find the proper $G$
such that (\ref{lagr1}-\ref{lagr2}) define a null surface by setting $u$ and 
$\phi $ constant. As we will show in this work, curves with singular points
in $E^{2}$ will induce null surfaces with caustics in $E^{3}$. Conversely,
we will also show that if the null surfaces have conjugate points then the
corresponding families of curves on $E^{2}$ have singularities.

In this reformulation, eq. (\ref{ec1}) for $Z$ is changed to an equation for 
$G$, of the same type, 
\begin{equation}
\frac{d^{3}G}{dp^{3}}=\widetilde{\Lambda }(p,G,G^{\prime },G^{\prime \prime
})  \label{ecG}
\end{equation}
and the metricity condition $M[\Lambda \,]=0$, which becomes singular at a
caustic point, yields an entirely well-behaved equation $\tilde{M}[
\widetilde{\Lambda }]=0$ whose solutions are the ``appropriate'' r.h.s. of (
\ref{ecG}). As we will see in an example, among the solutions of (\ref{ecG})
we obtain generating functions of caustics as listed in \cite{arnold}. This
result encounters an immediate application inside the 2+1 NSF and also in
the full theory because, to be in complete agreement with the standard GR
formulation, it always remained as an open problem to write the metricity
conditions of NSF in a form that explicitly shows the existence of the
singular solutions \cite{fkn}.

The work is organized as follows: in section II we first give an account of
Cartan's geometric construction obtained in \cite{cartan} starting from the
third-order equation (\ref{ec1}). This review is presented in modern
language since the original reference was written before modern differential
geometry was invented and it is very difficult to follow. ( It is worth
mentioning that one ``variation'' on Cartan's work has appeared in the
literature \cite{newman} but it is different from the original
construction.) We then give a brief review of the NSF in 3-dim and show that
this formalism is a particular case of Cartan's construction. In section III
we proceed to generalize the local analysis of the previous section and
write the regularized metricity condition $\tilde{M}[\widetilde{\Lambda }]=0$%
. We also find a relation between curves with caustics on $E^{2}$ and null
surfaces with conjugate points on $E^{3}$. A simple example shows how to
generate germs of caustics for the solutions to (\ref{ecG}). We conclude
this work with some comments on the possibility of attaching a similar
geometric structure to the original construction of the NSF in four
dimensions.

\section{Geometric approaches to a third-order ODE.}

In this section we present two geometric constructions that arise from a third 
order
ordinary differential equation 
\begin{equation}
u^{\prime \prime \prime }=\Lambda \,(\phi ,u,u^{\prime },u^{\prime \prime }),
\label{eq:thirdorder}
\end{equation}
where $(\phi ,u)$ are coordinates on the cylinder $S^{1}\times \mathbf{R}$.
As we will see below, it turns that one of these constructions is almost 
contained as a
particular case of the other.

\subsection{Cartan's construction.}

We first present Cartan's approach \cite{cartan,chern}, achieved by
interpreting the integral curves $u(\phi ,x^{a})$ as points  $x^{a}(a=1,2,3)$ of 
the three
dimensional solution space $E^{3}$, attaching to
it a Lorenztian metric $g_{ab}$ and giving the structure of a one
dimensional fiber bundle over $E^{3}$, with $\phi \in [0,2\pi )$ as the
coordinate of the fiber.  On the
base space we attach a ``null frame'' satisfying 
\begin{equation}
\mathbf{e}_{1}\cdot \mathbf{e}_{1}=\mathbf{e}_{3}\cdot \mathbf{e}_{3}=%
\mathbf{e}_{2}\cdot \mathbf{e}_{3}=0\quad \mbox{and}\quad \mathbf{e}%
_{2}\cdot \mathbf{e}_{2}+\mathbf{e}_{1}\cdot \mathbf{e}_{3}=0,  \label{frame}
\end{equation}
where $\mathbf{e}_{i}\cdot \mathbf{e}_{j}=g_{ab}\,\mathbf{e}_{i}{}^{a}%
\mathbf{e}_{j}{}^{b}=g_{ij}$. In terms of the dual basis  $\sigma
_{a}{}^{i}$ of the null frame, the metric can be written as 
\[
g_{ab}=g_{ij}\,\sigma _{a}{}^{i}\sigma _{b}{}^{i}=\sigma _{a}{}^{2}\otimes
\sigma _{b}{}^{2}-2\sigma _{a}{}^{(1}\otimes \sigma _{b}{}^{3)},
\]
where we have chosen $g_{13}=-1$. We call the bundle constructed in this way
as the bundle of null directions and we denote it by $\mathcal{N}(E^{3},g)$. 
Note that for fixed values of $\phi $, the three
functions 
\begin{equation}
(\theta ^{0},\theta ^{1},\theta ^{2})\equiv (u,\omega ,r)\equiv (u,u^{\prime
},u^{\prime \prime }),  \label{coord}
\end{equation}
can be taken as coordinates on the base space of the bundle. Thus, each
point of the bundle is locally described by $(\phi ,u,\omega ,r).$

Equation (\ref{eq:thirdorder}) yields the pfaffian system \cite{choquet} on $
\mathcal{N}(E^{3},g)$ which, written in terms of these coordinates reads
\begin{eqnarray}
\sigma ^{1} &=&\text{d}u-\omega \;\mbox{d}\phi,   \nonumber \\
\sigma ^{2} &=&\mbox{d}\omega -r\;\mbox{d}\phi,   \label{pfaffian} \\
\sigma ^{3} &=&\lambda \,(\mbox{d}r-\Lambda \,\,\mbox{d}\phi -\alpha
\,\sigma ^{1}-\beta \,\sigma ^{2}) \nonumber,
\end{eqnarray}
where $\alpha $, $\beta $ and $\lambda $ are functions to be determined. If
we choose the $\sigma _{a}{}^{i}$ to be the projections of (\ref{pfaffian})
to the base space, then the solution of (\ref{eq:thirdorder}) will give null
surfaces $S\in E^{3}$ by setting $u(\phi ,x^{a})=const.,\phi =const$.

To characterize the geometric structure defined from (\ref{eq:thirdorder}),
Cartan introduces a connection and a covariant exterior derivative as

\[
D\mathbf{e}_{i}=\omega _{i}\text{ }^{j}\text{ }\mathbf{e}_{j},
\]
where $D$ is the covariant exterior derivative with respect to this
connection (see\cite{Dieud,Kob,Traut}) and $\omega _{i}\text{ }^{j}\text{ }$ are 
the
connection one-forms. Furthermore, Cartan demands that the connection will be
compatible with the metric in the sense that a null frame remains null under
parallel transport, i.e.

\[
Dg_{ij}=2\bar{\omega}\,g_{ij},
\]
where $\bar{\omega}$ is a one-form on the bundle. Therefore,
\[
\omega _{ij}+\omega _{ji}=-2g_{ij}\,\bar{\omega}\quad \text{ where }\omega
_{ij}=\omega _{i}\text{ }^{k}g_{kj}.
\]
It follows from the above that

\begin{eqnarray*}
\omega _{11} &=&\,\omega _{33}=0,\text{ }\omega _{13}+\omega _{31}=-2\bar{%
\omega}, \\
\omega _{12} &=&-\omega _{21},\text{ }\omega _{23}=-\omega _{32},\text{ }%
\omega _{22}=\bar{\omega}.
\end{eqnarray*}
Thus, the connection is determined by four arbitrary one-forms, namely 
\[
\omega _{12},\,\omega _{23},\,\omega _{31}\;\;\text{and}\;\;\bar{\omega}.
\]

Note that $\sigma ^{i},\omega _{23}$ are linearly independent forms in $
\mathcal{N}(E^{3},g)$. This assertion can be understood from the geometrical
meaning that these forms have:

\begin{itemize}
\item  $\sigma ^{1}=\omega _{23}=0$ is the differential system for null
2-surfaces, since on this surface 
$D$ $\mathbf{e}_{3}=\omega _{13}$ $\mathbf{
e}_{3}$ and $D$ $\mathbf{e}_{2}=-\omega _{22}$ $\mathbf{e}_{2}+\omega _{12}$ 
$\mathbf{e}_{3}$.

\item  $\sigma ^{1}=\sigma ^{2}=\omega _{23}=0$ is the differential system
for null geodesic, since on this curve 
$D$ $\mathbf{e}_{3}=\omega _{13}$ $
\mathbf{e}_{3}$.

\item  $\sigma ^{1}=\sigma ^{2}=\sigma ^{3}=0$ is the differential system
for a point of $E^{3}$.

Note also that the vanishing of $\sigma ^{1}$ and $\omega _{23}$ is
equivalent to impose $u=$ const., $\phi =$ const. Thus, $\omega _{23}$ can
be chosen to be of the form $\omega _{23}=\mu ($d$\phi +\gamma \sigma ^{1})$
with $\gamma $ and $\mu $ being a priori arbitrary.
\end{itemize}

The idea is to write the non trivial connection one-forms in terms of the
basis $\sigma ^{i},\omega _{23}$ and then to impose certain conditions on
the torsion and curvature of the connection to uniquely determine the
functions $\alpha $, $\beta $, $\gamma $, $\lambda $ and $\mu $.

Using Cartan's structure equations 
\begin{eqnarray}
\Theta ^{i} &=&\mbox{d}\sigma ^{i}+\omega ^{i}{}_{j}\wedge \sigma ^{j},
\label{torsion} \\
\Omega ^{i}{}_{j} &=&\mbox{d}\omega ^{i}{}_{j}+\omega ^{i}{}_{k}\wedge
\omega ^{k}{}_{j},  \label{curvature}
\end{eqnarray}
and imposing 
\begin{equation}
\Theta ^{1}=\Theta ^{2}=0,\Theta ^{3}=A\,\sigma ^{1}\wedge \omega _{23}
\label{torsion cond}
\end{equation}
on the torsion two-forms, we find that 
\begin{eqnarray*}
\lambda  &=&\mu =1, \\
\alpha  &=&\frac{1}{3}\,\partial _{r}\Lambda ,\text{ } \\
\beta  &=&\alpha ^{2}-\frac{1}{2}\,\frac{\text{d}\alpha \,}{\text{d}\phi }+%
\frac{1}{2}\partial _{\omega }\Lambda,
\end{eqnarray*}
where

\begin{equation}
\frac{\text{d}F}{\text{d}\phi }(u,\omega ,r,\phi )=\partial _{\phi }F+{%
\omega }\,\partial _{u}F+r\,\partial _{{\omega }}F+\Lambda \,\,\partial _{r}F
\label{del}
\end{equation}
for any function $F(u,\omega ,r,\phi )$. Equation (\ref
{torsion cond}) has the following geometrical meaning: given a point in $E^{3}$ 
and two vectors, we construct a \text\it{geodesic parallelogram} from that point 
 (see \cite{goeck}); then in order to come back to the same point, we only need 
a 
translation in the null
direction $\mathbf{e}_{3}$. If the parallelogram is on the null surface
(when $\sigma ^{1}=0$ and $\omega _{23}=0$), no translation is needed. 

Furthermore, if we require 
\begin{equation}
\Omega _{23}=B\,\sigma ^{1}\wedge \sigma ^{2}+C\,\sigma ^{1}\wedge \sigma
^{3},  \label{curvaturecond}
\end{equation}
we find that 
\[
\gamma =\frac{1}{2}\partial _{r}\,\alpha .
\]
The geometrical meaning of (\ref{curvaturecond}) is that the above curvature
two-form should vanish when $\mathbf{e}_{3}$ is parallely transported around
a parallelogram with one of its sides being the null geodesic generated by $%
\mathbf{e}_{3}$ \cite{cartan}. 

Summarizing, conditions (\ref{torsion cond}) and (\ref{curvaturecond})
suffice to uniquely determine the non trivial components of the connection
one-form. They are given by

\begin{eqnarray}
\omega _{23} &=&\text{d}\phi +\gamma \text{ }\sigma ^{1},  \nonumber \\
\bar{\omega} &=&-\alpha \text{ d}\phi +\left( 2\frac{\text{d}\gamma \,}{%
\text{d}\phi }-\,\partial _{\omega }\alpha \right) \sigma ^{1}-2\text{ }%
\gamma \text{ }\sigma ^{2},  \label{connection} \\
\omega _{31} &=&\alpha \text{ d}\phi +\left( \frac{\text{d}\gamma \,}{\text{d%
}\phi }+\alpha \text{ }\gamma \right) \sigma ^{1}+\gamma \text{ }\sigma ^{2},
\nonumber \\
\omega _{12} &=&-\beta \text{ d}\phi +\left( \partial _{u}\alpha -\partial
_{\omega }\beta +3\beta \text{ }\gamma -\alpha \text{ }\partial _{r}\beta
\right) \sigma ^{1}+\left( 2\frac{\text{d}\gamma \,}{\text{d}\phi }%
-\,\partial _{\omega }\alpha \right) \sigma ^{2}-\gamma \text{ }\sigma ^{3}.
\nonumber
\end{eqnarray}
Using the above equations one determines the remaining coefficients for the
torsion and curvature in terms of $\Lambda $. In particular, the only non
trivial coefficient of the torsion is given by 

\[
A=\frac{1}{6}\,\frac{\text{d}^{2}\partial _{r}\Lambda \,}{\text{d}\phi ^{2}}-
\frac{1}{3}\,\partial _{r}\Lambda \,\,\frac{\text{d}\partial _{r}\Lambda \,}{
\text{d}\phi }-\frac{1}{2}\,\frac{\text{d}\partial _{\omega }\Lambda \,}{
\text{d}\phi }+\frac{2}{27}(\partial _{r}\Lambda \,)^{3}+\frac{1}{3}
\,\partial _{\omega }\Lambda \,\,\partial _{r}\Lambda \,+\partial
_{u}\Lambda \,.
\]

A connection compatible with a Lorenztian metric constructed in this way was
called by Cartan a normal metric connection. The main result of Cartan can
be stated as follows:

\begin{theorem}
{To each third order ordinary differential equation (up to diffeomorphism in 
$(\phi ,u)$) one can associate a null bundle with a unique normal metric
connection and viceversa, i.e. to each bundle of null directions with normal
metric connection one can associate a third order ordinary differential
equation up to diffeomorphism. }
\end{theorem}

One special class of Cartan's connection is particularly interesting to us.
If we impose $A=0$, the connection is torsion free and the Monge's equation 
\[
g_{ab}Y^{a}Y^{b}=0 
\]
is constant along the fiber, since 
\begin{equation}
\frac{\mbox{d}g_{ab}}{\mbox{d}\phi }=-\frac{2}{3}\partial _{r}\Lambda
\;g_{ab}+2\;A\,\sigma _{a}{}^{1}\otimes \sigma _{b}{}^{1}.
\label{derivmetric}
\end{equation}

Thus, the condition $A=0$ defines a unique conformal structure on the
solution space of (\ref{eq:thirdorder}). Moreover, $A=0$ is also the
condition that the third order ODE must satisfy so that the contact of two
neighboring integral curves can be given by a Monge's equation of the
second order between the parameters of those curves \cite{Wunchsmann}.

Note that even in the case that the connection is torsion free, the metric
so obtained depends on $\phi $, i.e. we have a monoparametric family of
conformally related metrics.

\subsection{NSF's construction.}

We now turn our attention to the NSF formulation, in which we have a
function $u=Z(\phi )$ satisfying a third-order ODE 
\begin{equation}
u^{\prime \prime \prime }\!=\Lambda \,(\phi ,u,u^{\prime },u^{\prime \prime
}\!),  \label{eq:ODE}
\end{equation}
with $\phi $ $\in $ $S^{1}$ and $\Lambda \,$ a smooth generic function. The
solutions to this equation are of the form $u=Z(x^{a};\phi )$, with $%
x^{a}(a=1,2,3)$ representing three constants of integration which define the
3-dim manifold of solutions $E^{3}$ (equivalent to $R^{3}$).

Note that the function $Z(x^a; \phi)$ plays a double role, namely:

\begin{itemize}
\item  For each fixed $x^{a}$ in $E^{3}$, $u=Z(x^{a};\phi )$ yields a curve $%
C_{x}$ on $E^{2}$ with coordinates $(u,\phi )$; these curves will be called
cuts.

\item  Fixing $(u,\phi )$ in $E^{2}$, the relation $u=Z(x^{a};\phi )$
defines now a surface $S_{(u,\phi )}$ living in $E^{3}$.
\end{itemize}

It is important to realize that the analysis is merely done at a local
level, so that the curve $C_{x}$ is certainly the graph of a function in $%
E^{2}$ and $S_{(u,\phi )}$ turns out to be a smooth surface in $E^{3}$.

The key assumption of NSF comes when we require that, for any value of $u$
and $\phi $, $S_{(u,\phi )}$ be indeed a null surface of some space-time
metric $g_{ab}(x^{a})$ to be attached to $E^{3}$. This condition implies
that for fixed values of $x^{a}$ and arbitrary values of $\phi $ the
gradient of $Z$ satisfies 
\begin{equation}
g^{ab}(x^{a}){\nabla }_{a}Z(x^{a};\phi ){\nabla }_{b}Z(x^{a};\phi )=0.
\label{null}
\end{equation}
Note that as the families of foliations intersect at a single but arbitrary
point $x^{a}$, the enveloping surface forms the light cone of the point $
x^{a}$. Thus, the parameter $\phi $ spans the circle of null directions.

The idea now is to consider (\ref{null}) as an algebraic equation from which
the five components of the conformal metric can be determined in terms of $%
\nabla _{a}Z(x^{a},\phi )$. Given an arbitrary function $Z$, the problem has
no solution since we have an infinite number of algebraic equations (one for
each value of $\phi $) for five unknowns. Therefore, we must impose
conditions on $Z(x^{a},\phi )$ so that a solution exists. The solution and
conditions are obtained by repeatedly operating $\frac{d}{d\phi }$ on~($\!\!$%
~\ref{null}). They are best expressed when written in the canonical
coordinate system $(u,\omega ,r)$ given in eq.(\ref{coord}). The final
expression for the metric components reads

\begin{equation}
g^{ij}=\Omega ^{2}\left( 
\begin{array}{ccc}
0 & 0 & 1 \\ 
0 & -1 & -\frac{1}{3}\partial _{r}\Lambda \, \\ 
1 & -\frac{1}{3}\partial _{r}\Lambda \, & -\frac{1}{3}\partial (\partial
_{r}\Lambda \,)+\frac{1}{9}(\partial _{r}\Lambda \,)^{2}+\partial 
_{\omega}\Lambda \,
\end{array}
\right),  \label{metric}
\end{equation}
where the conformal factor must satisfy the differential equation (see \cite
{2+1} for details) 
\[
\frac{d}{d\phi }\Omega =\frac{1}{3}\partial _{r}\Lambda \,\Omega , 
\]
so that the metric is independent of $\phi $. Note that the above equation
is invariant under $\Omega (x,\phi )\to \Omega ^{\prime }(x,\phi
)=f(x)\Omega (x,\phi )$ for an arbitrary $f(x)$. This freedom is a
consequence of the conformal invariance of the formulation.

The metricity condition is given by 
\[
M[\Lambda ]\equiv 2\left( \frac{{d(\partial _{r}\Lambda \,)}}{{d\phi }}
-\partial _{\omega}\Lambda \,-{\frac{2}{9}}(\partial _{r}\Lambda \,)^{2}\right)
\partial _{r}\Lambda \,-\frac{{d^{2}(\partial _{r}\Lambda \,)}}{{d\phi }^{2}}
+3\frac{{d(\partial _{\omega}\Lambda \,)}}{{d\phi }}-6\partial _{u}\Lambda
\,=0, 
\]
and constraints the available $\Lambda \,$s that must enter in the
differential equation (\ref{eq:ODE}), for only if $M[\Lambda \,]=0$ holds,
one is able to construct from the solutions $Z(x^{a};\phi )$ to the ODE a
metric according to (\ref{metric}) such that the level surfaces $S_{(u,\phi
)}$ of $Z$ are its characteristic surfaces. Note that this condition is the
one deduced by Cartan imposing the connection in the three dimensional
manifold $E^{3}$  to be torsion free ($A=0$).

Summarizing, one solves $M[\Lambda ]=0$, which is a partial differential
equation in the variables $(u,\omega ,r,\phi )$ and, denoting by $\Lambda
\,_{0}(u,\omega ,r,\phi )$ its solution and using the coordinates
definitions (\ref{coord}), equation (\ref{eq:ODE}) becomes 
\begin{equation}
Z^{\prime \prime \prime }\!=\Lambda \,_{0}(\phi ,Z,Z^{\prime },Z^{\prime
\prime }\!),
\end{equation}
which is the ODE whose solutions $Z(x^{a};\phi )$ allow for the construction
of a metric $g_{ab}$ on $E^{3}$ such that the level surfaces of $Z$ are its
null hypersurfaces.

\begin{remark}
It is clear from the above that the NSF construction is the special case of
the Cartan's geometric structure with vanishing torsion.
\end{remark}


\section{ Singularities in terms of smooth manifolds}


By assumption, both Cartan's and the NSF are local constructions in the
2-dim manifold. It follows from this that the cuts $C_{x}\subset E^{2}$ are
families of smooth curves and (as we will see below) that the level sets $Z={%
u_{0}}$ are smooth null surfaces of the induced metric in the 3-dim solution
space.

It also follows from the above assumption that these formalisms are not
capable to include the caustics that null surfaces necessarily possess \cite
{arnold, arnold1, arnold2} . Moreover, one might foresee that the families
of cuts that yield these null surfaces with caustics will also fail to be
smooth, developing self-intersections and singular points. Thus, in this
section, we are faced with the problem of generalizing the geometric
constructions presented in the previous section to include the description
of singularities of both, cuts and null hypersurfaces.

Our starting assumption is that the cuts develop caustics in the 2-dim
space. Technically this means that cuts are projections onto $E^{2}$ of a
(smooth) Legendrian submanifold and the caustics arise where the projection
map fails to be one to one. In a similar way as in \cite{ko-la-re}, we use a
generating function to describe a cut in the neighborhood of a caustic.
Using this function we will be able to see how $\Lambda $ diverges at a
caustic point. Since the original equation and the metricity condition
become useless around that point, we will obtain a regularized metricity
condition to select the class of generating functions which yield conformal
metrics on the solution space. We will also show that caustics come
inseparably paired in the Cartan and NSF constructions, in the sense that
the existence of caustics in the cut implies the same singular behavior in
the null surfaces of the manifold $E^{3}$ (and viceversa). The section ends
with an example of a generating function in a minkowskian space-time.

\subsection{The generating function}

The regular solutions of (\ref{eq:ODE}) can be used to define smooth curves
on the so called projective cotangent space of $E^{2}$ with local
coordinates $(u,\phi ,\pi )$, where $\pi $ is the momentum canonically
conjugated to $\phi $. The curves are given by

\begin{eqnarray}
u &=&Z(x^{a};p),  \nonumber \\
\phi &=&p,  \nonumber \\
\pi &=&\frac{dZ}{dp},  \nonumber
\end{eqnarray}
with $x^{a}$ parameters to be interpreted as coordinates in the solution
space. These smooth curves are called Legendrian submanifolds and the
projection of these submanifolds onto the space $(u,\phi )$ gives the local
description of the cuts of $E^{2}$. The above equations describe the
Legendrian submanifolds in the diffeomorphic region since the coordinate $%
\phi $ is used as a parameter to describe these curves.

In order to describe the cuts in regions containing caustic points, we
introduce a generating function of the form $G=G(x^{a};p)$. In this case the
Legendrian submanifold is given locally as

\begin{eqnarray}
u &=&G-p\text{ }G^{\prime },  \nonumber \\
\phi &=&-\text{ }G^{\prime },  \label{contact} \\
\pi &=&p,  \nonumber
\end{eqnarray}
where $\displaystyle G^{\prime }$ denotes the derivative of $G$ with respect
to $p$ holding $x^{a}$ fixed. The above equations locally describe smooth
curves on the projective cotangent space of $E^{2}$. Note that the
projection of these curves onto the space $(u,\phi )$ is not necessarily a
diffeomorphism, since $\phi $ fails to be injective in $p$ when $G^{\prime
\prime }$ vanishes. Therefore, this description includes caustic points in a
natural way. It is easy to see that $\Lambda $ diverges at the caustic
points. For this we analyze the behavior of the coordinates $(u,\omega
,r,\phi )$ and the function $\Lambda \,$ as we approach a caustic in $E^{2}$%
. These variables, expressed in terms of $G$ and $p$, become

\begin{eqnarray}
u &=&G-pG^{\prime },  \nonumber \\
\omega &=&\frac{du}{d\phi }=p,  \label{eq:coord} \\
r &=&\frac{d\omega }{d\phi }=-\left( G^{\prime \prime }\right) ^{-1}, 
\nonumber
\end{eqnarray}
and since $\displaystyle{\Lambda =\frac{d^{3}u}{d^{3}\phi }}$ we obtain 
\[
\Lambda =\frac{-G^{\prime \prime \prime }}{\left( G^{\prime \prime }\right)
^{3}}. 
\]

Since at a caustic point $\displaystyle G^{\prime \prime }{=0}$, we see from
equations (\ref{eq:coord}) that both $u$ and $\omega $ are bounded while $r$
and $\Lambda $ diverge at that point. One might argue that $G$ could be such
that $G^{\prime \prime \prime }$ also vanishes at a caustic point in such a
way that $\Lambda $ remains finite. However, smoothness requires $G$ to be
expandable as a polynomial around a caustic point. Thus $G^{\prime \prime
\prime }$ is always a polynomial of lower degree than $G^{\prime \prime }$
and the previous argument does not apply. It follows from these
considerations that $a)$ the third order ODE (\ref{eq:ODE}) will not be
defined around caustic points and $b)$ the coordinate system, metric
construction and metricity condition will not be regular on the 3-dim
solution space.

We look then for a regular third order ODE where the right hand side of this
equation satisfies a regularized metricity condition. For this we introduce
a new set of coordinates $(G,G^{\prime },G^{\prime \prime },p)$. In terms of
these coordinates we have 
\[
G^{\prime \prime \prime }=\tilde{\Lambda}(G,G^{\prime },G^{\prime \prime
},p)=-\text{ }(G^{\prime \prime })^{3}\Lambda (G-pG^{\prime },p,-(G^{\prime
\prime })^{-1},-G^{\prime }) 
\]
and the metricity condition $\tilde{M}[
\widetilde{\Lambda }]=0$ becomes

\begin{eqnarray}
\tilde{M}[
\widetilde{\Lambda }] &=&G^{\prime \prime }\left( 2(\tilde{\Lambda}%
_{p}+G^{\prime }\tilde{\Lambda}_{G})\tilde{\Lambda}_{G^{\prime \prime }}-3%
\tilde{\Lambda}\frac{d}{dp}\tilde{\Lambda}_{G^{\prime \prime }}-5\frac{d%
\tilde{\Lambda}}{dp}\tilde{\Lambda}_{G^{\prime \prime }}+2\tilde{\Lambda}(%
\tilde{\Lambda}_{G^{\prime \prime }})^{2}\right)  \nonumber \\
&&+\,G^{\prime \prime }\left( 3\frac{d^{2}\tilde{\Lambda}}{dp^{2}}-3\frac{d}{%
dp}(\tilde{\Lambda}_{p}+G^{\prime }\tilde{\Lambda}_{G})\right) +(G^{\prime
\prime })^{2}\left( 2\tilde{\Lambda}_{G^{\prime \prime }}\frac{d}{dp}\tilde{%
\Lambda}_{G^{\prime \prime }}-6\tilde{\Lambda}_{G}\right)  \label{eq:metreg}
\\
&&-\,(G^{\prime \prime })^{2}\left( \frac{4}{9}(\tilde{\Lambda}_{G^{\prime
\prime }})^{3}+\frac{d^{2}}{dp^{2}}\tilde{\Lambda}_{G^{\prime \prime
}}\right) +3\tilde{\Lambda}\left( (\tilde{\Lambda}_{p}+G^{\prime }\tilde{%
\Lambda}_{G})+\tilde{\Lambda}\tilde{\Lambda}_{G^{\prime \prime }}-\frac{d%
\tilde{\Lambda}}{dp}\right) =0,  \nonumber
\end{eqnarray}
where $\frac{dF}{dp}=F_{p}+G^{\prime }$ $F_{G}+G^{\prime \prime }$ $%
F_{G^{\prime }}+\tilde{\Lambda}$ $F_{G^{\prime \prime }}$ for any function $%
F(G,G^{\prime },G^{\prime \prime },p)$. Note that the above equation is
regular in a neighborhood of the caustic ($G^{\prime \prime }=0$).

Finally, to obtain null surfaces and study them near a caustic, we proceed
in a similar way as we did in the diffeormophic region, i.e. we first solve 
$\tilde{M}[
\widetilde{\Lambda }]=0$. Denoting by $\tilde{\Lambda}_{0}$ a particular
solution to this equation we then generate solutions of the ordinary
differential equation 
\begin{equation}
G^{\prime \prime \prime }=\tilde{\Lambda}_{0}(G,G^{\prime },G^{\prime \prime
},p),  \label{eq:ODEG}
\end{equation}
and construct families of curves with caustics on $E^{2}$.

\begin{remark}
Note that in the $(G,p)$ space the solutions to (\ref{eq:ODEG}) will
generate smooth curves. To construct the families of curves with caustics
one must use $G$ as the generating function of the contact transformation (%
\ref{contact}) and the null surfaces are given by the conditions $u$ =
const., $\phi $ = const.
\end{remark}

\begin{remark}
Note that the contact transformation (\ref{contact}) induces a coordinate
transformation (\ref{eq:coord}) which preserves the metric tensor defined on 
$E^{3}$. Thus, Cartan's theorem 2.1 is immediately generalized to include
coordinate and contact transformations on $E^{2}$.
\end{remark}

\subsection{Caustics on null surfaces and cuts}

\indent

In this subsection, we prove that the existence of caustic points in the cuts
yields the existence of caustic points in the null surfaces of $E^{3}$ and
viceversa. As it is known (\cite{hawking}), the divergence of a congruence
of null geodesics $\rho =-\nabla _{a}l^{a}$ ($l^{a}=g^{ab}\nabla _{a}Z$)
becomes infinite at a caustic. Therefore, to prove our assertions, we will
derive a relation between  $\rho $ of a congruence contained in the null
surface  and the scalar $Z^{\prime \prime }(x^{a},\phi )$ constructed from
the general solution $u=Z(x^{a},\phi )$ to (\ref{eq:ODE}) .

Let $l^{a}$ be the tangent vector to a geodesic with affine parameter $\tau $%
, then 
\begin{equation}
l^{a}=\frac{dx^{a}}{d\tau }=g^{ab}\theta ^{0}{}_{b}=\Omega ^{2}\theta
_{2}{}^{a},  \label{eq:la}
\end{equation}
where $\theta ^{\text{ }i}{}_{b}$ and $\theta _{j}{}^{a}$ are the form-basis
and its corresponding dual vector-basis associated with the canonical
coordinate system (\ref{coord}).

To this tangent vector we associate a triad $\{l^{a},m^{a},n^{a}\}$
parallely propagated along the geodesic satisfying 
\[
m^{a}m_{a}=1,\qquad n^{a}n_{a}=0,\qquad l^{a}m_{a}=0,\qquad
l^{a}n_{a}=-1\quad \mbox{and}\quad n^{a}m_{a}=0.
\]
In this frame the metric tensor and the divergence become 
\[
g_{ab}=m_{a}m_{b}-2\text{ }n_{(a}l_{b)}\qquad \mbox{and}\qquad \rho
=m^{a}m^{b}\nabla _{a}l_{b}
\]
respectively. Since $\theta _{1}{}^{a}$ and $\theta _{2}{}^{a}$ are
coordinate vectors they satisfy 
\begin{equation}
\theta _{2}{}^{a}\nabla _{a}\theta _{1}{}^{b}=\theta _{1}{}^{a}\nabla
_{a}\theta _{2}{}^{b}.  \label{eq:gamma}
\end{equation}
Expresing  the geodesic deviation vector $\theta _{1}{}^{a}$ 
in terms of the triad

\begin{equation}  \label{eq:devi}
\theta_1{}^a=\xi m^a+\alpha l^a,
\end{equation}
 in (\ref{eq:gamma}) gives 

\[
l^{a}\nabla _{a}\xi =\xi m^{a}m^{b}\nabla _{a}l_{b}=\xi \,\rho 
\]
or equivalently 
\begin{equation}
\rho =\xi ^{-1}\frac{d\xi }{d\tau }.  \label{eq:rho}
\end{equation}
On the other hand we can easily derive a differential equation for the
divergence by considering the geodesic deviation equation as follows: 
\[
m_{c}l^{a}\nabla _{a}(\theta _{1}{}^{b}\nabla _{b}l^{c})=m_{c}\theta
_{1}{}^{a}\nabla _{a}(l^{b}\nabla _{b}l^{c})+\xi
R_{abd}{}^{c}l^{a}m^{b}l^{d}m_{c}.
\]
Since $l^{a}$ is a tangent vector of a null geodesic and the triad is
parallely propagated along it, using (\ref{eq:devi}) we obtain 
\[
m_{c}l^{a}\nabla _{a}(\xi m^{b}\nabla _{b}l^{c})=\frac{d\xi }{d\tau }\rho
+\xi \frac{d\rho }{d\tau }=\xi \Phi ,
\]
with $\Phi =R_{abd}{}^{c}l^{a}m^{b}l^{d}m_{c}$. Finally, from (\ref{eq:rho}
), we see that $\rho $ must satisfy the differential equation 
\begin{equation}
\frac{d\rho }{d\tau }+\rho ^{2}=\Phi.  \label{eq:diff}
\end{equation}

To prove our claims we start with the general solution $u=Z(x^{a},\phi )$ to
(\ref{eq:ODE}), and we assume that the cut  $u=Z(x_{0}^{a},\phi )$ has a
caustic at $\phi =\phi _{0}$. As we have seen in the previous subsection,
this is equivalent to say that the function $Z^{\prime \prime
}(x_{0}^{a},\phi )$ diverges at $\phi _{0}$. To show that there is a caustic
in the null surface defined by $Z(x^{a},\phi _{0})=Z(x_{0}^{a},\phi
_{0})=u_{0}$, we must prove that $\displaystyle{\lim_{r\to \infty }\rho
=\infty }$, with $r=Z^{\prime \prime }(x^{a},\phi _{0})$. 

From (\ref{eq:la}) it follows that
\[
\frac{d}{d\tau }=\Omega ^{2}\frac{\partial }{\partial r}=-\Omega ^{2}s^{2}%
\frac{\partial }{\partial s},
\]
with the coordinate $s=r^{-1}$. As $\frac{d}{d\tau }$ and $\frac{\partial }{%
\partial s}$ are regular operators near $s=0$, the factor $\Omega ^{2}s^{2}$
must be a non zero smooth function of $s$. Hence $\Omega =\mathcal{O}(s^{-1})
$, meaning that the conformal factor $\Omega $ diverges when we approach a
caustic in the cut.

Finally, noting that $g_{11}=g_{ab}\theta _{1}{}^{a}\theta _{1}{}^{b}$, we
express $\xi $ in terms of the conformal factor in the following manner 
\[
g_{11}=\Omega ^{-2}=g_{ab}(\xi m^{a}+\alpha l^{a})(\xi m^{b}+\alpha
l^{b})=\xi ^{2},
\]
therefore 
\[
\rho =\frac{d}{d\tau }\ln \Omega =-\Omega ^{2}s^{2}\frac{\partial }{\partial
s}\ln \Omega
\]
which yields 
\[
\rho =\mathcal{O}(s^{-1}).
\]
We conclude that when $s$ goes to zero (in a singular point of the cut) $
\rho $ diverges, i.e. we approach a caustic of the null surface.

\vspace{0.2in} It remains now to prove that a caustic in the null surface
leads to the same singular behaviour in the cut. Suppose that the null
surface is defined by $u_{0}=Z(x^{a},\phi _{0})$, then near a caustic point $
x_{0}^{a}$ in the surface the solution of the differential equation (\ref
{eq:diff}) in the flat case yields

\[
\rho =\frac{1}{\tau -\tau _{0}},
\]
where the singular point $x_{0}^{a}$ corresponds to $\tau =\tau _{0}$. By
means of (\ref{eq:rho}) we find $\xi =\Omega ^{-1}=\tau -\tau _{0}$, thus
the conformal factor $\Omega $ diverges when we approach a caustic in the
null surface and since $d\tau =\Omega ^{-2}dr$, $r$ diverges (for $\tau $ is
the affine parameter of the curve and so $d\tau $ is bounded) at $x_{0}^{a}$.

On the other hand we know that $u=Z(x_{0}^{a},\phi )$ is a solution of (\ref
{eq:ODE}), then at the singular point $x_{0}^{a}$ the function $r=Z^{\prime
\prime }(x_{0}^{a},\phi _{0})$ diverges. Thus, the cut $u=Z(x_{0}^{a},\phi )$
also has a caustic point at $\phi =\phi _{0}$.

\begin{remark}
Note that the spaces $E^{2}$ and $E^{3}$, used to describe the cuts and null 
surfaces
respectively, are not related in any way. The only tool that was used to prove 
the above
results relating the cuts with null surfaces is the starting ODE (\ref{eq:ODE}).
This is in contrast with similar results obtained in 3 and 4 dimensions where 
the
reciprocity theorem for null congruences and the assumption that $E^{2}$ is a
hypersurface of $E^{3}$ is explicitly used in the proof \cite{2+1}.      

\end{remark}


\subsection{ An example: caustics in flat space}


An arbitrary constant function is clearly a solution of the metricity
condition $\tilde{M}[\widetilde{\Lambda }]=0$, which yields a polynomial of
third degree in $p$ for $G$. Therefore we choose $\widetilde{\Lambda }=-1$
and (\ref{eq:ODEG}) yields 
\[
G(p,x^{a})=-\frac{p^{3}}{3}+\frac{1}{2}\,X^{1}{p}^{2}+X^{2}p+X^{3}.
\]

With this generating function the Legendre submanifold is given
parametrically by 
\begin{eqnarray}  \label{eq:u}
u&=&\frac{2}{3}\,{p}^{3}-\frac{1}{2}\,X^1{p}^{2}+X^2, \\
\phi&=&p^2-X^1p-X^2, \\
p&=&p.
\end{eqnarray}

\begin{figure}[htp]
\begin{center}
\includegraphics[angle=-90,width=6cm]{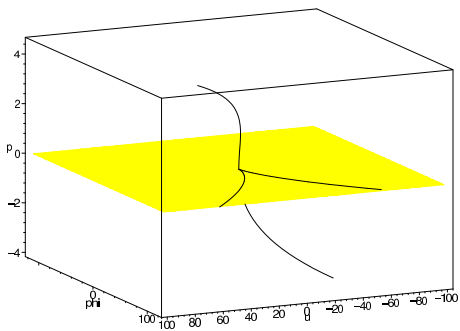} \hspace{.3cm}
\end{center}
\caption{{\protect\small Legendre submanifold and the corresponding cut for $%
X^1=X^2=X^3=0$.}}
\label{fig1}
\end{figure}

The projection of this submanifold onto $E^{2}$, $(u,\phi ,p)\to (u,\phi )$,
is the cut and possesses a cusp-like caustic when $p=X^{1}/2$ (see figure 
\ref{fig1}).

\vspace{.2in}

Since $\phi$ is not an injective function of $p$, there are two values of $p$
for each $\phi$, namely 
\begin{equation}  \label{eq:pphi}
p(\phi)=\frac{1}{2}X^1\pm\frac{1}{2}\sqrt{(X^1)^2+4\phi+4X^2}.
\end{equation}

Inserting this expression into equation (\ref{eq:u}) we obtain the
corresponding space-time coordinates 
\begin{eqnarray}
u &=&\frac{1}{12}\,\left( {(X^{1})}^{2}+4\,\phi +4\,X^{2}\right) ^{3/2}+%
\frac{1}{12}\,{(X^{1})}^{3}+\frac{1}{2}\,X^{1}\,\phi +\frac{1}{2}%
\,X^{1}\,X^{2}+X^{3}  \nonumber \\
\omega &=&\frac{1}{2}X^{1}+\frac{1}{2}\,\sqrt{{(X^{1})}^{2}+4\,\phi +4\,X^{2}%
} \\
r &=&{\frac{1}{\sqrt{{(X^{1})}^{2}+4\,\phi +4\,X^{2}}}}.  \nonumber
\end{eqnarray}

It is clear from (\ref{eq:pphi}) that $\sqrt{(X^1)^2+4\phi+4X^2}$ becomes
null in the caustic, hence $r$ diverges while $u$ and $\omega$ remain
bounded at this point.

Figure 2 shows a null surface and null geodesics on it for $(u,\phi)=(0,0)$,
which end tangent to the caustic curve.
\begin{figure}[hp]
\begin{center}
\includegraphics[angle=-90,width=6cm]{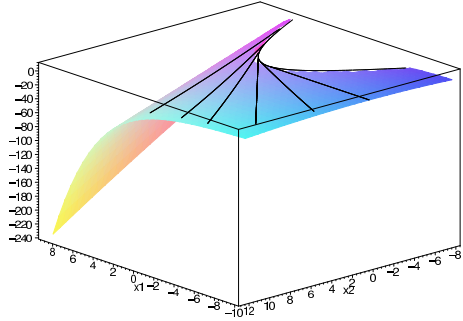} \hspace{.3cm}
\end{center}
\caption{{\protect\small Null surface for $(u,\phi)=(0,0)$. }}
\end{figure}
The above generating function gives $\Lambda=-r^3$, which yields a flat metric
for the conformal factor $\Omega=r$. 

\section{Conclusions}

\indent

We have shown in this work how the 3-dim NSF can be recast in terms of a
well known mathematical frame, like Cartan's geometrical theory of
differential equations. In this manner, the metricity condition of NSF
becomes a simple geometric imposition within the Cartan's framework, namely
the vanishing of the torsion of the connection in the space $E^{4}$.

Moreover, both constructions have been extended from their local scopes to
the non-differentiable regions, in order to account for the singular
behavior that null surfaces necessarily possess. As result of this extension,

\begin{itemize}
\item  {a generalized version of the metricity condition was obtained, whose
solutions yield null foliations of a 2+1 dimensional manifold with caustics
and other singularities that the local construction of NSF by definition is
not capable to describe.}

\item  {the singularities of cuts and null surfaces were shown to be closely
related in the sense that the singular behavior in one of them induces a
similar behaviour in the other.}
\end{itemize}

At this point, one can think of applying the same ideas of this work to the
original 4-dim version of NSF \cite{fkn, kn}. This would imply a change from 
ordinary
differential equations over partial ones, since the variable $Z$ now depends
on two coordinates $(\alpha ,\beta )$ on $S^2$ and is subject to the following 
system
of PDEs:
\begin{equation}
Z_{\alpha \alpha }=\Lambda (\alpha ,\beta ,Z,Z_{\alpha },Z_{\beta
},Z_{\alpha \beta }),  \nonumber
\end{equation}
\begin{equation}
Z_{\beta \beta }=\Upsilon (\alpha ,\beta ,Z,Z_{\alpha },Z_{\beta },Z_{\alpha
\beta }),  \nonumber
\end{equation}
where $Z_{\alpha },Z_{\beta }$ are the partial derivatives of $Z$ with
respect to the coordinates. Since the above system of equations is
integrable, its solution space can be parametrized by four constants $x^{a}$%
, which locally define the 4-dim solution space $M$. Thus, one would be able
in principle to construct an $S^{2}$ bundle over $M$ and attach a similar
set of geometric structures that have been presented in this paper.

If this program can be done, the NSF could be described within a well
established context of differential equations; one would be able to give a
geometrical interpretation to the metricity conditions (possibly in terms of
requirements analog to the vanishing of the torsion). 

\subsection*{Acknowledgments}

This research has been partially supported by AIT, CONICET, CONICOR, and
UNC. We thank Paul Tod for enlightening conversation.


\begin{thebibliography}{99}
\bibitem{cartan}  E. Cartan, \textit{La geometr\'{\i }a de las Ecuaciones
Diferenciales de Tercer Orden}, Rev. Mat. Hispano-Americana, \textbf{IV},
1-31, (1941).

\bibitem{cartan2}  E. Cartan, \textit{Les espaces g\'{e}n\'{e}ralis\'{e}s e
l'int\'{e}gration de certaines classes d'\'{e}quations diff\'{e}rentielles},
Academie des Sciences du Paris, 1689-1693 (1938).

\bibitem{chern}  S. Chern, \textit{The geometry of the differential equation 
$y^{\prime \prime \prime }=F(x,y,y^{\prime },y^{\prime \prime })$},
Collected work, t. \textbf{I}, 385-438 (1939).

 \bibitem{2+1}D. Forni, M. Iriondo y C. Kozameh, {\it Null surfaces formulation 
in 3-d}. {\bf Vol 41}, 5517-5534,  Journal of 
Mathematical Physic (2000).  

\bibitem{hawking}  S. Hawking and G.F.R. Ellis, \textit{The large scale
structure of space-time}, Cambridge University Press, Cambridge (1973).



\bibitem{arnold}  V. I. Arnold, \textit{Mathematical Methods of Classical
Mechanics}, Springer-Verlag, Berlin Heidelberg, NY (1984).

\bibitem{fkn}  S. Frittelli, C. N. Kozameh and E. T. Newman, \textit{GR via
characteristic surfaces}, J. Math. Phys., \textbf{36}, 4984-5004 (1995).


\bibitem{newman} S. Frittelli, E. T. Newman and  C. N. Kozameh  
\textit{Differential Geometry from Differential Equations}, submitted to 
Comm. Math. Phys., preprint (2000).

\bibitem{choquet}  Y. Choquet-Bruhat, C. De Witt-Morette and M.
Dillard-Bleick, \textit{Analysis, manifolds and physics}, North-Holland
Publishing Company (1977).

\bibitem{Dieud}  Dieudonn\'{e}, \textit{Elementos de analisis}, Revert\'{e},
Espa\~{n}a, vol IV (1983).

\bibitem{Kob}  Kobayashi and Nomizu, \textit{Foundations of Differential
Geometry}, Jhon Wiley and Sons, New York, vol I (1963)).

\bibitem{Traut}  Andrzej Trautman. \textit{Differential Geometry for
Physicists} Stony Brook Lecture, Bibliopolis (1984).

\bibitem{goeck}   M. G\"{o}ckeler and T. Sch\"{u}cker. \textit{Differential
geometry, gauge theories and gravity. }Cambridge Monographs on Mathematical
Physics, Cambridge University Press (1987).

\bibitem{Wunchsmann} K. Wunschmann, \textit{Ueber Beruhrungsbedingungen bei
Integralkurven von Differentialgleichungen,} Inaug.Dissert., Leipzig, Teubner
\bibitem{arnold1}  V. I. Arnold and Novikov \textit{Dynamical systems}, 
\textbf{IV}, Springer-Verlag, Berlin Heidelberg, NY, (1992).

\bibitem{arnold2}  V. I. Arnold, \textit{Singularities of Caustics and Wave
fronts}, Kluwer, Dordrecht (1990).


\bibitem{ko-la-re}  C. N. Kozameh, P. W. Lamberti, O. A. Reula, \textit{
Global aspects of light cone cuts}, \textit{J.Math.Phys}, \textbf{32},
(1991).





\bibitem{kn}  C. N. Kozameh and E. T. Newman, \textit{Theory of light cone
cuts of null infinity} J. Math. Phys., \textbf{24}, 2481-2489 (1983).






\end{thebibliography}
\end{document}